\documentclass[12pt]{article}

\usepackage{amsmath,amssymb,graphicx} %use drftcite in draftmode
\usepackage{epsf}
\usepackage{pstricks}
\usepackage{cite}
\newcommand{\beq}{\begin{eqnarray}}% can be used as {equation} or  {eqnarray}
\newcommand{\eeq}{\end{eqnarray}}

\newcommand{\centeron}[2]{{\setbox0=\hbox{#1}\setbox1=\hbox{#2}\ifdim

\wd1>\wd0\kern.5\wd1\kern-.5\wd0\fi
\copy0

\kern-.5\wd0\kern-.5\wd1\copy1\ifdim\wd0>\wd1
                                       \kern.5\wd0\kern-.5\wd1\fi}}
\newcommand{\ltap}{\>\centeron{\raise.35ex\hbox{$<$}}
                               {\lower.65ex\hbox{$\sim$}}\>}
\newcommand{\gtap}{\>\centeron{\raise.35ex\hbox{$>$}}
                               {\lower.65ex\hbox{$\sim$}}\>}

\newcommand\ZZ{\hbox{\zfont Z\kern-.4emZ}}
\font\zfont = cmss10 %scaled \magstep1

\def\gappeq{\mathrel{ \rlap{\raise.5ex\hbox{$>$}}
                      {\lower.5ex\hbox{$\sim$}}  } }
\def\lappeq{\mathrel{ \rlap{\raise.5ex\hbox{$<$}}
                      {\lower.5ex\hbox{$\sim$}}  } }
\begin{document}
\begin{titlepage}
\begin{flushright}
{\tt hep-ph/0703031
}
\end{flushright}

\vskip.5cm
\begin{center}
{\LARGE BRIDGE:\\{\bf B}ranching {\bf R}atio {\bf I}nquiry/{\bf
D}ecay {\bf G}enerated {\bf E}vents
 \vspace{.2cm}}

\vskip.1cm
\end{center}
\vskip0.2cm

\begin{center}
{\bf Patrick Meade$^a$ and Matthew Reece$^b$}
\end{center}
\vskip 8pt

\begin{center}
{\it $^a$Jefferson Physical Laboratory\\
Harvard University, Cambridge, MA 02138, USA} \\
\vspace*{0.3cm}
{\it $^b$Institute for High Energy Phenomenology\\
Newman Laboratory of Elementary Particle Physics\\
Cornell University, Ithaca, NY 14853, USA } \\
\vspace*{0.3cm}
\end{center}

\vglue 0.3truecm

\begin{abstract}
\vskip 3pt \noindent We present the manual for the program
BRIDGE: Branching Ratio Inquiry/Decay Generated Events. The
program is designed to operate with arbitrary models defined
within matrix element generators, so that one can simulate events
with small final-state multiplicities, decay them with BRIDGE, and
then pass them to showering and hadronization programs.
BRI can automatically calculate widths of
two and three body decays. DGE can decay unstable
particles in any Les Houches formatted event file.
DGE is useful for the generation of event files with long decay
chains, replacing large matrix elements
by small matrix elements followed by sequences of decays.
BRIDGE is currently designed to work with
the MadGraph/MadEvent programs for implementing and
simulating new physics models. In particular, it can operate with
the MadGraph implementation of the MSSM. In this manual we describe how
to use BRIDGE, and present a number of sample results to demonstrate
its accuracy.
\end{abstract}

\end{titlepage}

\newpage

\renewcommand{\thefootnote}{(\arabic{footnote})}

%%%%%%%%%%%%%%%%%%%%%%%%%%%%%%%%%%%%%%%%%%%%%%%%%%%%%%
%%%%%%%%%%%%%%%%%%%%%%%%%%%%%%%%%%%%%%%%%%%%%%%%%%%%%%
\section{Introduction}
\setcounter{equation}{0}
\setcounter{footnote}{0}
%%%%%%%%%%%%%%%%%%%%%%%%%%%%%%%%%%%%%%%%%%%%%%%%%%%%%%
%%%%%%%%%%%%%%%%%%%%%%%%%%%%%%%%%%%%%%%%%%%%%%%%%%%%%%

In recent years the workhorses of Monte Carlo event generation,
parton showering, and hadronization, Pythia~\cite{Pythia} and
Herwig~\cite{Herwig}, have begun to offload some of their duties
to other programs. It is now possible to define new models,
automatically generate matrix elements, and simulate parton-level
events using a variety of available packages (among them
CompHEP/CalcHEP~\cite{comphep,calchep}, MadGraph~\cite{Madgraph},
AcerMC~\cite{AcerMC}, Amegic++~\cite{Amegic}, and Grace~\cite{Grace}).
The events
simulated in these packages can then be passed to the next step in
the chain, a parton-shower code, using a standardized ``Les
Houches Accord" file format~\cite{EventFiles,Skands:2003cj}.

These developments (and others; we apologize for the many software
packages we cannot mention in this brief summary) have made the
task of implementing a model of new physics and studying its
collider signatures far easier than it has been in the past. On
the other hand, in our experience there have been two bottlenecks
in the process, which demanded a general solution. In BRIDGE we
provide (at least a useful first approximation to) that solution.

The first is that in order to accurately calculate matrix
elements, one wants to know the widths of new unstable particles
in the model. There exist specialized tools (e.g.
HDECAY~\cite{HDECAY}, SDECAY~\cite{SDECAY}, and
SUSY-HIT~\cite{SUSYHIT}) for particular models. There is also a
general calculator in CompHEP~\cite{comphep} (and its
spin-off, CalcHEP~\cite{calchep}), but to use it with
another event generator one must translate the entire model into
CompHEP's formats.
It seemed desirable to have a general,
independent calculator of tree-level decay widths. BRIDGE can
currently read MadGraph-style definitions of particles and
interactions and use them to calculate widths, but it can easily
be extended to read input in other formats as well.

The main advantage of BRIDGE over CompHEP/CalcHEP as
a width calculator is that it adds decay functionality, which overcomes
the second bottleneck. The problem is simulating events that involve long decay
chains, which are familiar in the MSSM context (e.g. decays of
squarks or gluinos) but also arise in many other models of more
recent vintage. In an ideal world we would all have enough
computing power to integrate a full $2 \rightarrow N$ phase space
where $N$ could be as large as 8 or 10. Unfortunately, practical
limitations on computer power make it desirable to be able to
simulate $2 \rightarrow 2$ or $2 \rightarrow 4$ processes and then
decay the unstable final-state particles. This factorizes the
matrix elements, and one loses information both to narrow-width
approximations and to the loss of interference effects, but for
most purposes it seems to be a fairly good approximation and it
greatly speeds up computations. However, there has not been a
general code that can decay unstable particles for arbitrary
models and keep the original helicity structure of the vertex
(i.e., go beyond a flat phase space approximation).\footnote{Pythia
can do arbitrary decays, but will only use flat phase space (it
does not know the amplitude).
It must be supplied with the quantum numbers and decay tables
of the new particles. Both CalcHEP and BRIDGE can be
used with Pythia in this way.}
BRIDGE provides that code.

BRIDGE operates in two pieces, ``BRI" (Branching Ratio Inquiry,
though it really calculates widths) and ``DGE" (Decay Generated
Events). The BRI stage uses Vegas integration~\cite{LepageVegas}
of the phase space for the decay to calculate a width. The amplitudes
are computed using the HELAS libraries~\cite{HELAS}. The DGE
stage uses the stored grids from BRI to choose random points in
the phase space to use to decay actual events. BRIDGE owes much to
Fabio Maltoni's ``DECAY" code included in MadGraph, which plays a
similar role but only for Standard Model particles.

In Section~\ref{sec:program} we will describe the structure of the
program and how to use it. In Section~\ref{sec:examples} we will
show a variety of examples of calculations performed with BRIDGE
and compared to known answers or to simulations with full matrix
elements.

%%%%%%%%%%%%%%%%%%%%%%%%%%%%%%%%%%%%%%%%%%%%%%%%%%%%%%
%%%%%%%%%%%%%%%%%%%%%%%%%%%%%%%%%%%%%%%%%%%%%%%%%%%%%%
\section{Program Structure and Use}\label{sec:program}
\setcounter{equation}{0} \setcounter{footnote}{0}
%%%%%%%%%%%%%%%%%%%%%%%%%%%%%%%%%%%%%%%%%%%%%%%%%%%%%%
%%%%%%%%%%%%%%%%%%%%%%%%%%%%%%%%%%%%%%%%%%%%%%%%%%%%%%
In this section we will discuss how BRIDGE should be ``installed"
and how to run BRI and DGE.  We will also briefly discuss how the program
works, for the benefit of the user who might wish to read or modify the
source code. The implementation of BRIDGE is in C++ and is driven by
several classes which allow BRIDGE to easily accommodate any
model.  In addition to the basic classes there are several
input/output functions that are driven by the user's interactive
choices in the executables {\tt runBRI.exe} or {\tt runDGE.exe}.

The program BRIDGE comes packaged inside a tarball {\tt
BRIDGEvX.XX.tar.gz}\footnote{This manual is current for the
version BRIDGEv2.00.} that should be unzipped and untarred in the
main Madgraph directory (e.g. {\tt MG\_ME\_V4.1.xx/}). This
relative location is used to find the HELAS libraries and the
model file directories in MG4\footnote{Should the user wish to use
a different directory structure: the location of the HELAS
libraries can be changed easily in the makefile, and the path to
the models directory is in the cpp files associated with DGE and
BRI.}. After untarring BRIDGE you will have a directory {\tt
MG\_ME\_V4.1.xx/BRIDGE} that includes the subdirectories {\tt
source, input} and {\tt results}. All files relevant to generating
the BRI and DGE executables are located within the {\tt source}
directory, and the {\tt input} directory has several model files
associated with the SM and MSSM for examples.  To generate the
executables associated with BRIDGE simply run {\tt
make}\footnote{It is assumed that you have already made Madgraph
so there exists the compiled library for HELAS.} from the main
BRIDGE directory. The makefile will create the two executables
previously mentioned, as well as {\tt runBRIsusy.exe} or {\tt
runDGEsusy.exe} which will be discussed in Section~\ref{sec:MSSM}.
Additionally beyond the described functions, from v2.0 there are
additional batch modes for all executables that will be described
in Section~\ref{sec:BATCH}.

\subsection{BRI}
BRI is designed such that given a definition of a model, including
numerical values for couplings and masses, all two and three body
partial widths can be calculated within that model at tree level.
By default only the standard renormalizable vertices are included,
via the HELAS libraries. In addition it is also possible to include
loop decays if implemented by the user.
An example will be given in Section~\ref{sec:Hgg}.

When {\tt runBRI.exe} is run, it first parses the model files for
all particles and interactions in a
given model.  It then will parse the files associated with the
numerical couplings and masses.  From this definition of a model
it will construct all possible kinematically allowed two and three
body decays for a given particle.  At this point the user
provides information on which particles they want BRI to calculate
the appropriate widths for.  BRI then calculates the widths by
integration, using the Vegas algorithm~\cite{LepageVegas}.
BRI also stores the grids generated by Vegas so that they can
be reused when decaying particles with DGE.

The actual implementation that BRI uses is based upon the use of
MG4 model definitions and the HELAS library for setting up the
matrix elements.  The Madgraph model definitions involve minimal
syntax. There is a {\tt particles.dat} which simply lists a particle name
and its relevant properties (e.g. antiparticle name, name of a parameter
that stores its mass, etc.), and an {\tt interactions.dat} which contains
a list of vertices and a name for the relevant coupling. We refer to the
Madgraph documentation~\cite{Madgraph} for details.

The real work in defining the model is to provide the numerical values
for all of the couplings in the model. BRI can read a simple text file
containing a list of coupling names (which should match that specified
in {\tt interactions.dat}) followed by up to four numbers. If one number
is specified, it is taken to be a real parameter. If two are specified,
it is a complex parameter; the first number is the real part, the second
the imaginary part. Finally, couplings involving fermions
require four real parameters
to specify: the first two are the real and imaginary parts of the left-handed
coupling, and the latter two are the real and imaginary parts of the
right-handed coupling. For example, here are a few lines from the
default Standard Model file {\tt SMparams.txt}:
\begin{verbatim}
WMASS 80.419
G 1.228
GG -1.228 0.0 -1.228 0
gzzhh       0.27638     0.00000
gal       0.31345      0.00000       0.31345      0.00000
\end{verbatim}
Note that extra whitespace is ignored (but each coupling should be on its own line),
and the coupling name is case-insensitive.

There is unfortunately not a universal format for a ``model" of physics
beyond the standard model, but the MG4 collaboration has provided
a relatively easy-to-use and well-defined format in their example models
and their ``usrmod" which is designed to easily
incorporate user implemented models. We have elected to choose their
formats for use in BRIDGE, but in principle our code can operate with
other formats if the relevant input/output code is added. Additionally, as
will be further discussed in Section~\ref{Sec:runbri}, the
Madgraph collaboration has provided a utility to numerically
generate the list of couplings for any user implemented model through
their {\tt usrmod} in the format that BRIDGE reads.

\subsubsection{Running BRI}\label{Sec:runbri}

The {\tt runBRI.exe} executable has two modes of operation. The
first is designed to seamlessly interface with the Madgraph 4
usrmod. The user must specify the model directory name, and BRI
will find the appropriate files there. Alternatively, the user can
specify all the necessary input files
as well as the location where the results will be placed.

We first describe the working of the more interactive mode.
There are three files that are requested by {\tt runBRI.exe}.
The first two are the definitions of a model
in Madgraph form, i.e. files in the format of {\tt particles.dat} and
{\tt interactions.dat}.  The third is the file that
includes the numerical values of the couplings and masses in the
model, an example of which is given for the SM in {\tt
BRIDGE/input/SMparams.txt}\footnote{The file name is irrelevant,
so long as it has the right format. Default values of the file
names are shown at the prompt for the input files.}.
The next step is to determine the list of particles for which widths
are to be computed and grids to be made. The user
is then presented with a list of all particles available for calculation,
and prompted to choose either a subset of these particles or to
calculate decays of all of them. By default the widths are calculated
only for a particle, not for its antiparticle, although you can also
ask for the antiparticle decay when specifying the list of particles
to decay. When DGE decays event files, it needs decay tables and
grids for the antiparticles as well. For that purpose we provide a script,
{\tt antigrids.pl}, which will symbolically link the tables for
antiparticles to those for particles. (We explain more about the
operation of this script in the first paragraph of Section~\ref{Sec:DGE}.)
Once the particles to be
decayed are determined, the user is prompted to specify input
parameters for the Vegas integration.  The user is asked for a
seed for the random number generator for Vegas (the default is the
time) and additionally the number of calls (the default is
50000) and of iterations (the default is 5). Calculating the
decays is the most computationally intensive part of BRIDGE, so if
you are calculating a large number of particles, be careful in your
choice of number of calls!  At this point BRI loops over the
particles requested and calculates all two and three body decays
for the given particle. For each particle a file {\tt "particle
name"\_decays.table} is created with a table of branching ratios
for the particle's decays.  Additionally for every decay mode of
the particle there is a corresponding {\tt ".grid"} file from
Vegas created with a filename that has the parent particle and the
corresponding daughters in the filename.  These files will be
used by DGE when decaying the particles.  Also, a
decay table of all particles is written to a file of your choosing
in Les Houches format~\cite{Skands:2003cj}.

The alternate run mode of BRI uses a specified Madgraph usrmod
directory. The tools in usrmod can take the work in generating
a file with numerical couplings out of the user's hands.
When using this mode of BRI the
user is required to have implemented the model in Madgraph
exactly as the ``usrmod" instructions explain.
Additionally one will have to have run {\tt make couplings} in the
usrmod directory and then run the executable {\tt couplings} which
makes a file {\tt couplings\_check.txt}.  This file will have
numerically evaluated all couplings for the Madgraph model and
thus the user is not required to have made a numerical couplings
file as they would have in the interactive mode.  Additionally the
user is required to have edited the file {\tt param\_card.dat} to include
the numerical values of the masses of the new particles.
One should note that this requires
the user to do nothing more than they would have for any
implementation in Madgraph.  When executing BRI with the Madgraph
usrmod all one specifies is the name of the Madgraph model
directory, and then the execution is identical to the other mode of
running BRI.  The only other difference in the Madgraph running
mode is that the output has an additional feature. For any
particles whose decays are calculated by BRI in this mode, the
widths for the particles are updated in the {\tt param\_card.dat}
for the model, so that after executing {\tt runBRI.exe} in this
mode one can run Madgraph with the correct widths automatically
implemented.

\subsection{DGE}\label{Sec:DGE}

We will now describe the working of DGE.  The program DGE is based
primarily upon the original decay program included in Madgraph,
written by Fabio Maltoni.  The original decay program for Madgraph
was hard wired to the Standard Model and each particular decay for
an SM particle was implemented separately. The main difference in
DGE is that it is ``model-independent" and automatically will
search for all possible decays for the particles of a given model.
To run DGE it is first necessary to have executed {\tt runBRI.exe}
before {\tt runDGE.exe} so as to have generated all the necessary
Vegas grids that DGE will use. If you created grids for a particle
but not its antiparticle (the default), you should run {\tt antigrids.pl}.
The script takes two command
line arguments. The first is the path to the {\tt particles.dat} so that the
script can learn which particle name is associated to which antiparticle
name. The second is the path to the results directory where the grids
are located. The script will then copy each existing decay table to
one appropriate for the antiparticle, provided this does not already exist.
It will also symbolically link the grids for the antiparticle
 to those for the particle.

DGE will again prompt you for the input files, just as BRI did. Then it
will ask for an input events file. (Currently, BRIDGE has only been
tested on MadEvent output files. If you need to use it with the output of
some other event generator, you either need to make sure the format
matches, edit the BRIDGE code yourself, or contact us for assistance.)
You will then be asked for a name
for the output events file, and the directory where the BRI grids are
stored. Also, you will again be prompted for a random number seed
which defaults to the current time.

DGE will give you three different options for what to do with the input file:
\begin{verbatim}
  Choose a mode:
    1. Decay a particular particle.
    2. Decay down to a set of final-state particles.
    3. Decay using a specified set of decay modes.
\end{verbatim}

In mode 1, you specify some particle to decay. DGE
 will read the decay table and grids generated
by BRI and use them to decay each instance of the particle chosen
 in the input file, choosing decay
modes randomly according to their branching ratio.

In mode 2, you specify a set of particles at which to stop. (For instance,
if you are decaying MSSM events, you might want to list the SM particles
and the LSP; then the output of DGE could be passed on through Pythia.)
DGE will prompt you either to enter
such a list at the command line (one particle at a time, typing ``END" to stop)
or to read the list from a file. If you choose the former option, you are also
asked if you would like to save the list of particles you have entered to a
file for future use.

In mode 3, you specify a set of decay modes to use. For each event, DGE
will look for particles with decay modes in the specified list. If it finds more than
one possible decay mode for a particle, it will choose randomly among the
specified modes according to their relative branching ratios. The output
events are weighted according to the fraction of each particles' decay mode
represented in the event. An example should make this clearer. Suppose I
simulate $W^+ W^-$ events and ask DGE to decay the $W^+$ using the
leptonic decay modes (about 1/3 of the overall branching fraction)
and the $W^-$ using the hadronic decay modes. Then each event will have
its original weight multiplied by $2/9$. The reason for this is that if you want
to combine events from multiple files decayed in different ways, this weighting
might prove useful. Again, you can either read in the modes from
a file (each line of which is simply a list like {\tt "w+ u d\~{}"} of mother particle
followed by daughter particles), or enter the modes at the command line,
and in the latter case you will be prompted about saving to a file.

Note that in all of these modes of running DGE, when there are multiple decay
modes for a particle DGE will choose among them based on their branching
ratios. If for some reason you do {\em not} wish to do this, one easy workaround
is to edit by hand the decay table created by BRI to reflect the ratio of events you
wish to you get.

\subsubsection{Helicities and angular distributions}\label{sec:DGEangles}

If DGE is running in either of the modes that decay multiple
particles at a time, it will attempt to preserve angular
information from the matrix element. To do so, it will boost any
given particle to the frame of the mother particle highest up the
chain. All helicities are reported in this frame, appearing in the
\texttt{SPINUP} column of the LHA format. For instance, if one
decays (for a sufficiently heavy Higgs) $h \rightarrow W^+ W^-$
followed by $W^+ \rightarrow \tau^+ \nu_\tau$, the \texttt{SPINUP}
output for the $\tau^+$, $\nu_\tau$, and $W^+$ will be the
helicity of the particle in the rest frame of the $h$ (and each
decay step will be computed with definite helicity in that frame).
The reason for this is that in passing between various frames,
helicities of massive particles can flip. Thus if successive decay
steps are performed in frames with relative boosts, we tend to
lose more of the angular information than we do by performing all
decays in one fixed frame. We will discuss an example of this in
Section~\ref{sec:Wpolarize}.

The key approximation that is being made in this code is that at
each stage of a decay process, each particle has a definite
helicity in the frame of the mother highest up the chain. Thus
certain interference effects will be absent in the output of
BRIDGE. For applications that rely on a complete understanding of
full spin correlations, one should run a program like MadGraph
that can generate the final state with the full matrix element.
For many-body final states this can be quite computationally
intensive. On the other hand, BRIDGE will preserve some of the
spin effects, as for each decay it uses the proper helicity
amplitudes. A more detailed study of the errors induced by
this approximation in various processes would be very interesting.

\subsection{Batch Modes}\label{sec:BATCH}

For all the executables generated for BRIDGE there are
corresponding command lines modes in addition to the default
interactive modes.  These are provided so that if it is desired to
scan over a parameter space or decay many files BRIDGE can be
incorporated inside another program.  The command line options are
as follows:
\begin{itemize}
\item runBRI has two modes depending on whether you are using MG
usrmod:
\begin{itemize}
\item MG usrmod:\\
{\tt runBRI y [model name] [blist particles elist] [vegas seed]
[calls] }\\{\tt [iterations]}

\item generic run: \\ {\tt runBRI n [particle file] [parameter
file] [interactionsfile] [directory for grids] [slha file] [blist
particles elist] [vegas seed] [calls]}\\{\tt [iterations] }
\end{itemize}
\item runDGE has two modes depending on whether you are using MG
usrmod:
\begin{itemize}
\item MG usrmod:\\
{\tt runDGE y [modelname] [input lhe file] [output lhe file]
[vegas seed] [specify final state particles or decay modes:2/3]
[file with decay list]}

\item generic run: \\ {\tt runDGE n [particle file] [parameter
file] [interactions file] [input lhe file] [output lhe file]
[directory for grids] [vegas seed] [specify final state particles
or decay modes:2/3] [file with decay list]}
\end{itemize}
\item runBRIsusy:
\begin{itemize}
\item {\tt runBRIsusy [particle file] [parameter file]
[interactions file] [directory for grids] [slha file] [blist
particles elist] [vegas seed] [calls]}\\ {\tt [iterations]}
\end{itemize}
\item runDGEsusy:
\begin{itemize}
\item {\tt runDGEsusy [particle file] [parameter file]
[interactions file] [input lhe file] [output lhe file] [directory
for grids] [vegas seed] [specify final state particles or decay
modes:2/3] [file with decay list]}
\end{itemize}
\end{itemize}

%%%%%%%%%%%%%%%%%%%%%%%%%%%%%%%%%%%%%%%%%%%%%%%%%%%%%%
%%%%%%%%%%%%%%%%%%%%%%%%%%%%%%%%%%%%%%%%%%%%%%%%%%%%%%
\section{Examples}\label{sec:examples}
\setcounter{equation}{0} \setcounter{footnote}{0}
%%%%%%%%%%%%%%%%%%%%%%%%%%%%%%%%%%%%%%%%%%%%%%%%%%%%%%
%%%%%%%%%%%%%%%%%%%%%%%%%%%%%%%%%%%%%%%%%%%%%%%%%%%%%%

\subsection{Top Decays: Basic Distributions}\label{sec:topdec}

\begin{figure}[h]
\begin{center}
\includegraphics[angle=270,width=11cm]{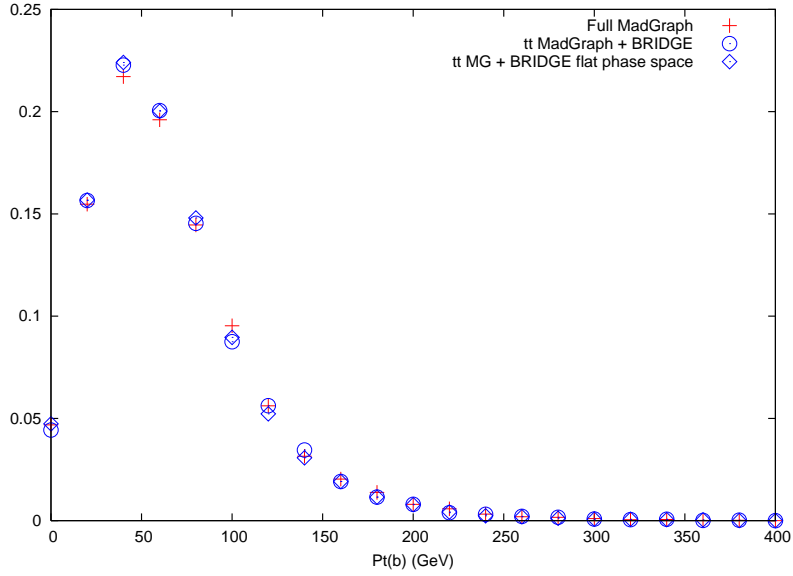}
\end{center}
\caption{We have generated $t\bar{t}$ events in MadGraph and
decayed them with BRIDGE, and also generated $e^+ \nu_e b e^-
\bar{\nu}_e \bar{b}$ events in MadGraph. Here we plot the $p_T$
histogram for the $b$ quark in the decayed events versus the full
matrix element. In this and other figures, the histograms are
normalized to have the same area. In this figure we also show
$t\bar{t}$ events from MadGraph decayed with BRIDGE with
the amplitude set to 1, so that the decay is governed by the phase
space volume.} \label{fig:bptcompare}
\end{figure}

\begin{figure}[!h]
\begin{center}
\includegraphics[angle=270,width=11cm]{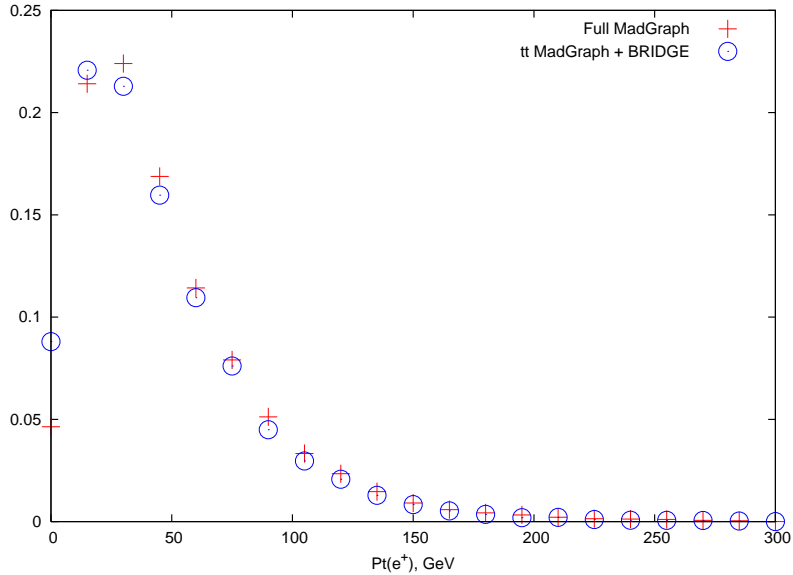}
\end{center}
\caption{The $p_T$ histogram for the $e^+$ in the decayed events versus the full matrix element.}
\label{fig:eptcompare}
\end{figure}

\begin{figure}[!h]
\begin{center}
\includegraphics[angle=270,width=11cm]{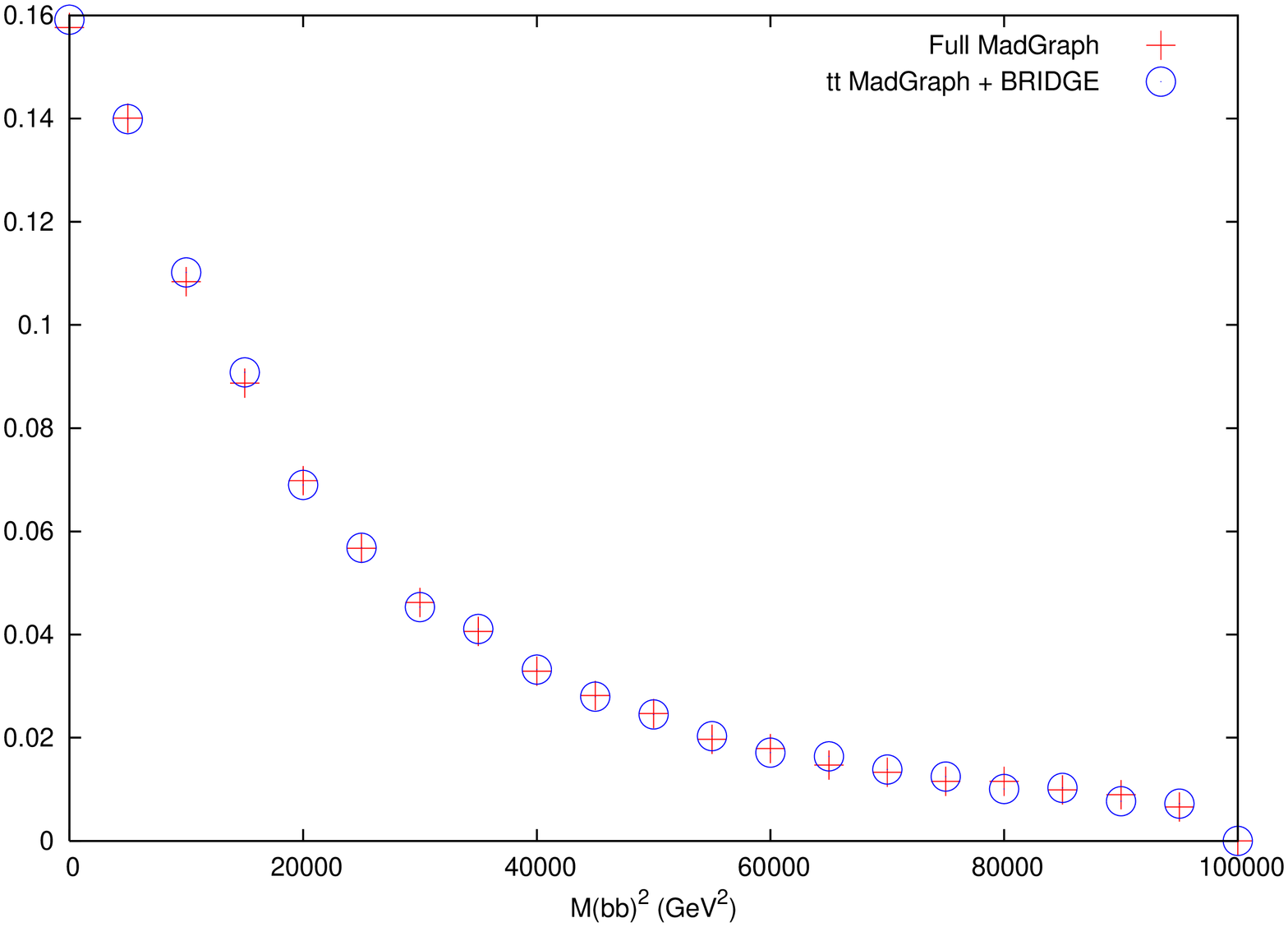}
\end{center}
\caption{The $M^2(b\bar{b})$ histogram in the decayed events versus the full matrix element.}
\label{fig:mbbcompare}
\end{figure}

\begin{figure}[!h]
\begin{center}
\includegraphics[angle=270,width=11cm]{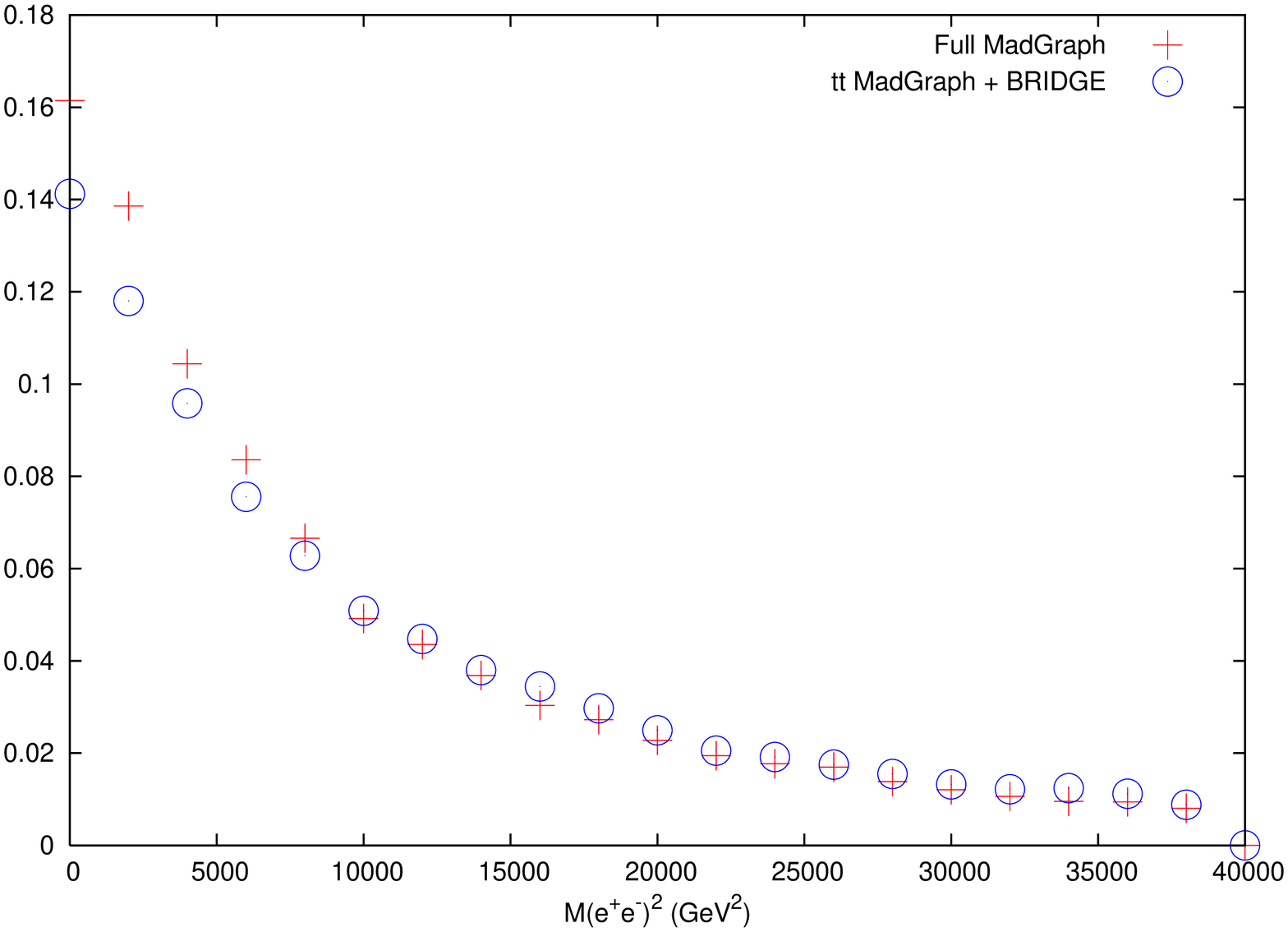}
\end{center}
\caption{The $M^2(e^+e^-)$ histogram in the decayed events versus the full matrix element.}
\label{fig:meecompare}
\end{figure}

As a first test of DGE, we present transverse momentum
distributions for the decay chain $t \rightarrow W^+ b \rightarrow
e^+ \nu_e b$, where the tops are chosen from $t\bar{t}$ events. We
generated $t\bar{t}$ events with MadGraph and decayed them in
BRIDGE, and also generated a set of $e^+\nu_e b e^- \bar{\nu}_e
\bar{b}$ events directly in MadGraph (demanding that the relevant
diagrams contain a $t\bar{t}$). The $p_T$ distribution for the $b$
quark is shown in Figure~\ref{fig:bptcompare}, and for the $e^+$
in Figure~\ref{fig:eptcompare}. The distributions agree reasonably
well. For the $p_T$ of the $b$ we also show the distribution computed
with a flat phase space approximation for the decay (i.e., we set
the amplitude to $1$ independent of the momenta). In this case,
the flat phase space agrees quite well. We will see a later example
for which the structure of the amplitude is important, and is captured
by BRIDGE, but a flat phase space amplitude is a poor approximation.

Next, we plot a more complicated quantity that involves information
from both sides of the event: the invariant mass squared
$M^2_{b\bar{b}}$, in Figure~\ref{fig:mbbcompare}, and
$M^2_{e^+e^-}$ in Figure~\ref{fig:meecompare}. There is a
discrepancy in $M^2_{e^+e^-}$, where BRIDGE seems to underestimate
the number of events with a very low invariant mass for the $e^+
e^-$ pair. Because invariant masses involve not only the momenta
but the angular distance between the particles, this suggests that
there might be some errors in angular distributions that involve
correlations between opposite sides of an event. Still, the
discrepancy is not huge (BRIDGE is about 13\% low for the first
bin).

Angular correlations can carry information about the helicity
structure of various couplings. We'll now turn to a specific
example and show that BRIDGE clearly distinguishes left- and
right-handed chiral gauge couplings.

\subsection{$W$ Polarization In Top Decays}\label{sec:Wpolarize}

\begin{figure}[!h]
\begin{center}
\includegraphics[angle=270,width=15cm]{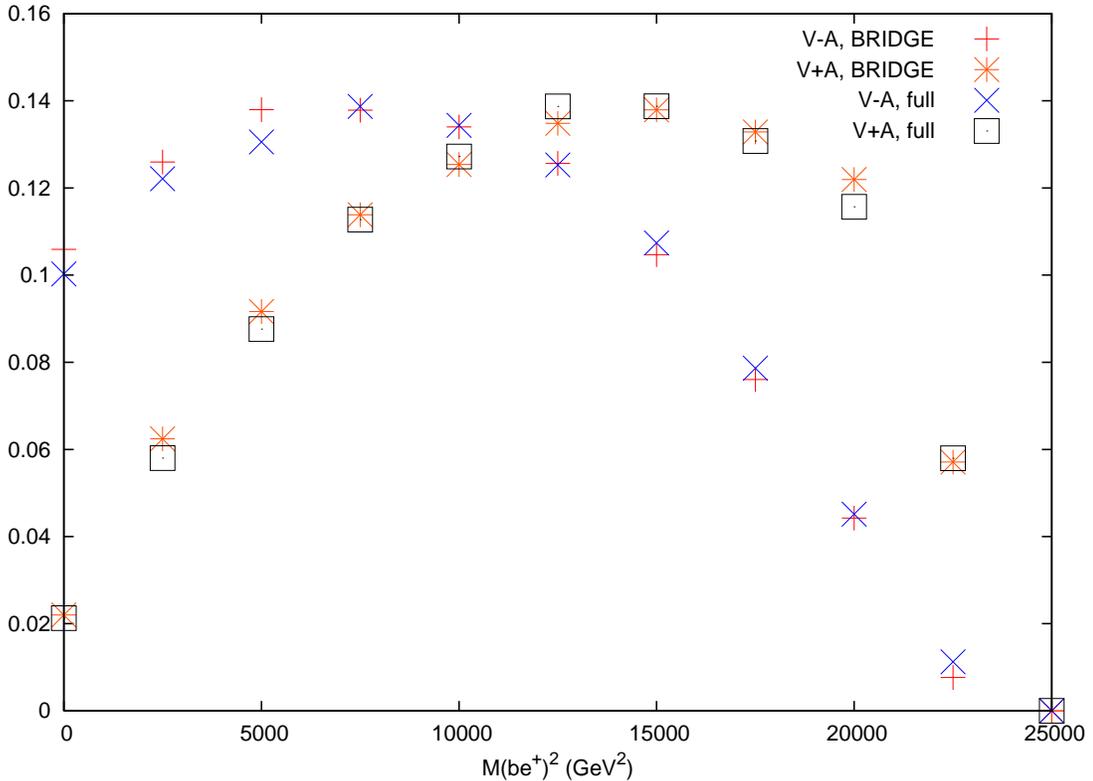}
\end{center}
\caption{Histogram of $M_{e^+ b}$ computed from MadGraph
$t\bar{t}$ events decayed with BRIDGE compared to MadGraph
$e^+\nu_e b \bar{t}$ events. In both cases the $V-A$ and $V+A$
structures for the $W^+\bar{t}b$ vertex are compared.}
\label{fig:DGEvsMG}
\end{figure}

Let's consider a simple example that illustrates the success of
BRIDGE in reproducing the proper helicity structure of decays: in
decays of the top quark, $t\rightarrow W^+ b$, we expect 70\% of
the $W$ bosons to be longitudinally polarized, 30\% to be
left-circularly polarized, and very few to be right-circularly
polarized. (The fraction of longitudinally polarized $W$s can be
computed as $\frac{x}{x+2}$ where $x = \frac{m_t^2}{m_W^2}\approx
4.7$.)

$W$ polarization is experimentally studied as a probe of the $V-A$
structure of the $tWb$ vertex. In this case one examines the decay
chain $t \rightarrow W^+ b \rightarrow \ell^+ \nu_\ell b$, where
$\ell = e,\mu$. The invariant mass $M_{\ell b}$ is then sensitive
to the angular structure of the decay. This has been studied
experimentally by the CDF Collaboration~\cite{Wpolarization}.

As a test of BRIDGE, we have simulated $t\bar{t}$ events with
MadGraph and then used DGE to decay $t \rightarrow W^+ b$, $W^+
\rightarrow e^+ \nu_e$. We have then produced a separate set of
decayed events for which we have altered the structure of the
$tWb$ vertex to be $V+A$. In each case we have plotted the
invariant mass $M_{e^+ b}$. We have produced the same plots with a
MadGraph simulation of production of $e^+ \nu_e b \bar{t}$
(requiring the $e^+ \nu_e b$ to come from a top), which will
maintain the full angular correlations. The results are compared
in Figure~\ref{fig:DGEvsMG}. The output of DGE is rather close to
the full MadGraph answer, and clearly shows the distinction
between $V-A$ and $V+A$.

\begin{figure}[!h]
\begin{center}
\includegraphics[angle=270,width=\textwidth]{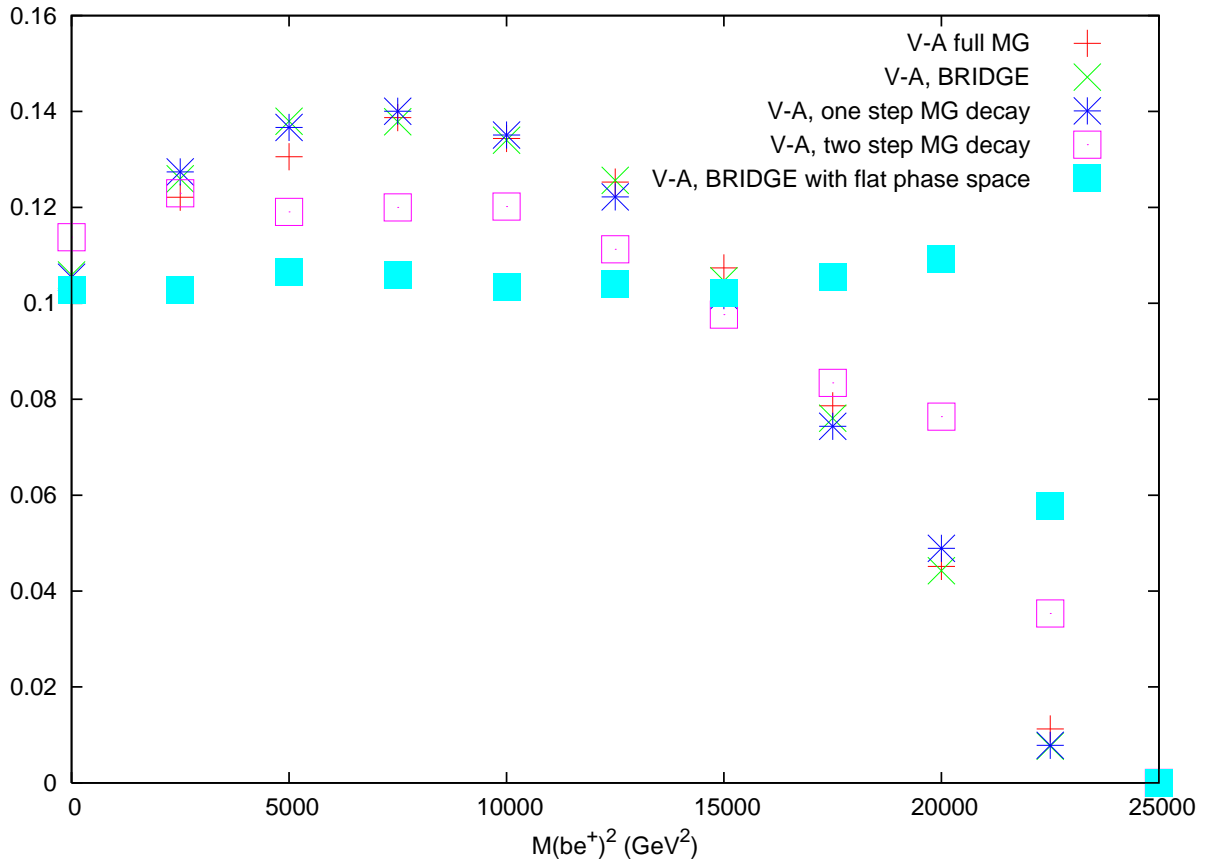}
\end{center}
\caption{Histogram of $M_{e^+ b}$ computed from MadGraph
$t\bar{t}$ events decayed with BRIDGE compared to the same events
decayed with MadGraph's ``decay" program in two steps ($t
\rightarrow W^+b$, $W^+ \rightarrow e^+ \nu_e$) or in one step,
and to full MadGraph $e^+\nu_e b \bar{t}$ events. The final curve
is BRIDGE with the amplitude simply replaced by $1$, to obtain
a distribution depending only on phase space volume. Only the $V-A$
structure for the $W^+\bar{t}b$ vertex is shown.}
\label{fig:DGEvsMGdec}
\end{figure}

As a comparison, we show in Figure~\ref{fig:DGEvsMGdec} the same
two curves for the $V-A$ structure only, as well as three additional
curves. The third curve comes from using the MadGraph decay code
to decay the same simulated $t\bar{t}$ events that we decayed with
DGE, using two sequential decay steps ($t \rightarrow W^+ b$
followed by $W^+ \rightarrow e^+ \nu_e$). The fourth curve uses
the MadGraph decay code to decay the same events, in only one step
($t \rightarrow e^+ \nu_e b$). For the one-step decay, MadGraph
builds the full amplitude and does a three-body phase space
integral instead of doing two successive decays. The final curve
shows the decay done in DGE but with the amplitude replaced by a
constant, so that the decay depends only on phase space volume.
From the figure,
it is apparent that DGE and the one-step MadGraph decay both model
the full MadGraph result accurately, whereas the two-step MadGraph
decay is less accurate. The two-step MadGraph decay is still significantly
more accurate than a naive flat phase space decay, however.
DGE and the two-step MadGraph decay are
performing nearly identical operations, but DGE is doing both
decay steps in the rest frame of the top. On the other hand,
MadGraph is doing decays in the center-of-mass frame of the
collision. As a result, in the frame MadGraph works in, many of
the $W^+$ bosons will be right-handed, and some of the information
about the angular structure is lost. The choice of frame that we
make in DGE allows for a more accurate result. (To see this, one
can generate $t\bar{t}$ events just above threshold in MadGraph.
On these events, the MadGraph two-step decay curve matches the
others quite well.) As we mentioned in
Section~\ref{sec:DGEangles}, we do not yet know precisely under
what conditions DGE gives reliable angular distributions and in
what conditions it loses important correlations. It would be very interesting
to study this further. For now, this example
shows that DGE can accurately reproduce some nontrivial
angular effects. We will see another example in Section~\ref{sec:MSSM}.

\subsection{MSSM}\label{sec:MSSM}

We briefly here discuss how the MSSM is designed to be used with
BRIDGE.  We provide two executables {\tt runBRIsusy.exe} and {\tt
runDGEsusy.exe} that are specifically designed for the MSSM. We
note however that this in no way means it is necessary to generate
new versions of the BRI and DGE executables for a generic model.
As discussed in Section~\ref{sec:program} any new model
implemented in the framework of the Madgraph usrmod can be
accommodated regardless of complexity. However, since the MSSM is
well defined in its couplings and there exists a standard SLHA
interface to spectrum calculators {\tt runBRIsusy.exe} and {\tt
runDGEsusy.exe} were designed to specifically work with this
format.  In the future the SUSY versions of runDGE and runBRI may
be reincorporated into the non-SUSY runDGE and runBRI, but for now
we will explain the existing interface.

As alluded to, the only main difference between the SUSY and
non-SUSY versions of BRIDGE is the input format.  As discussed in
Section~\ref{sec:program} the model is defined by four files, {\tt
particles.dat}, {\tt interactions.dat},  {\tt
couplings\_check.txt}, and {\tt param\_card.dat}.  The use of the
couplings and param card files are what defines the numerical
values of the masses and couplings in a generic usrmod file.
However, in the context of the MSSM there is a specific format for
defining the model parameters \cite{Skands:2003cj} and the
couplings of the model are well defined.  For this reason instead
of having the user only interface couplings through their
numerical values as in the usrmod version of input, the couplings
are defined separately in a file {\tt SUSYpara.cpp} and read
directly from {\tt param\_card.dat} through the SLHA read routines
in {\tt SLHArw.cpp}.  The coupling definitions found in {\tt
SUSYpara.cpp} are based upon those written originally for the
SMadgraph project~\cite{smadgraph}, that have since been
incorporated into Madgraph v4\footnote{We stress here that the
MSSM as defined in Madgraph/Smadgraph does not include all
interactions and the SUSY BRIDGE version is only as complete as
the assumptions in~\cite{smadgraph}.}. If one wanted to modify the
format of the MSSM couplings beyond the original assumptions
implemented in Smadgraph, the files {\tt SUSYpara.cpp} and {\tt
SLHArw.cpp} are all that are necessary to be modified.

The actual parameters used from the SLHA formatted input file are
those found in the blocks corresponding to mixing matrices for the
various supersymmetric particles, masses, SM inputs, A terms,
Yukawa couplings and Higgs parameters.  Additionally if available
{\tt BLOCK GAUGE} is used to define the SM gauge couplings evolved
to the scale specified by the spectrum calculator.  BRIDGE does
not run the SM couplings so {\tt BLOCK GAUGE} is used to define
couplings at a higher scale if available, if not the default
values at $m_Z$ are used.

The output of {\tt runBRIsusy.exe} is in the form of SLHA
formatted decay tables.  These decay tables can be then used with
any program that can handle SLHA formatted input. {\tt
runDGEsusy.exe} is used in the same way as {\tt runDGE.exe}. Given
the common definition of the MSSM, DGE can also be used to decay
Les Houches formatted event files created by other matrix element
generators for the MSSM.\footnote{The only discrepancies that can
arise from using DGE in this way are due to a difference in
definition of the MSSM from Smadgraph to another matrix element
generator.}.

\subsubsection{SLHA Decay Table Comparison}

In this section we demonstrate the numerical accuracy of using BRI
with the MSSM.  We will compare BRI against the well tested
SUSY-HIT program\cite{SUSYHIT}, which is the continuation of the
SDECAY program\cite{SDECAY} combined with HDECAY\cite{HDECAY}.  We
will for simplicity(lack of imagination) use the point SPS-1a, the
SLHA formatted spectrum card can be downloaded from the SUSY-HIT
webpage.  There are a few factors that influence the results of
this comparison.  First SUSY-HIT includes the effects of loops on
decays, whereas by default BRIDGE does not.  Therefore certain
decays for which SUSY-HIT includes loop corrections would be
expected to differ slightly.  Additionally decays that do not
exist in the MSSM at tree level are included in SUSY-HIT and would
have to be added separately into BRIDGE as will be discussed in
Section~\ref{sec:Hgg}.  Additionally any effects from running of
the couplings that are included in SUSY-HIT beyond the definition
of the couplings at the given scale in {\tt BLOCK GAUGE} will be
unaccounted for in BRIDGE.  Given that the full SLHA decay tables
for the MSSM would add many pages to this manual we will only
present as an example the decay tables for $\chi_2^{+}$ and
$\tilde{d}_L$ in Tables~\ref{tab:chidecay} and \ref{tab:dldecays}.

\begin{table}[ht]
\centering
\begin{tabular}{|l|c|c|}
\hline
& BRI & SUSY-HIT \\
$\chi_2^{+}$ Decays & $\Gamma_{\mathrm{BRI}}=$2.57908720E+00 & $\Gamma_{\mathrm{SUSY-HIT}}=$2.51618431E+00\\
 \hline
BR($\chi_2^{+}$ $\rightarrow$ $\chi_1^{+}$ Z )&     2.36511304E-01  & 2.40213561E-01  \\
BR($\chi_2^{+}$ $\rightarrow$ $\chi_1^{0}$ $W^{+}$ )&     6.55162205E-02  & 6.48085350E-02  \\
BR($\chi_2^{+}$ $\rightarrow$ $\chi_2^{0}$ $W^{+}$ )&     2.83642506E-01  & 2.86434649E-01  \\
BR($\chi_2^{+}$ $\rightarrow$ $\chi_1^{+}$ h )&     1.66651066E-01  & 1.70182080E-01  \\
BR($\chi_2^{+}$ $\rightarrow$ $\nu_e \tilde{e}^{+}$ )&     5.35091797E-02  & 5.44450726E-02  \\
BR($\chi_2^{+}$ $\rightarrow$ $e^{+} \tilde{\nu}_e$ )&     2.07052458E-02  & 2.10374477E-02  \\
BR($\chi_2^{+}$ $\rightarrow$ $\nu_\mu \tilde{\mu}^{+}$ )&     5.32258593E-02  & 5.44450726E-02  \\
BR($\chi_2^{+}$ $\rightarrow$ $\mu^{+} \tilde{\nu}_\mu$ )&     2.07805107E-02  & 2.10374477E-02 \\
BR($\chi_2^{+}$ $\rightarrow$ $\nu_\tau \tilde{\tau}_{1}^{+}$ )&     2.64808582E-03  & 1.86804725E-04  \\
BR($\chi_2^{+}$ $\rightarrow$ $\nu_\tau \tilde{\tau}_{2}^{+}$ )&     6.24909217E-02  & 5.89508529E-02  \\
BR($\chi_2^{+}$ $\rightarrow$ $\tau^{+} \tilde{\nu}_\tau$ )&     3.43191004E-02  & 2.82584766E-02 \\
\hline
\end{tabular}
\caption{$\chi_2^{+}$ decays in the MSSM for SPS-1a, calculated in
both BRIDGE and SUSY-HIT. The decay widths listed are in GeV.}
\label{tab:chidecay}
\end{table}

\begin{table}[ht]
\centering
\begin{tabular}{|l|c|c|}
\hline
& BRI & SUSY-HIT \\
$\tilde{d}_L$ Decays  & $\Gamma_{\mathrm{BRI}}=$5.25084508E+00 & $\Gamma_{\mathrm{SUSY-HIT}}=$5.25959716E+00\\
\hline
 BR($\tilde{d}_L$ $\rightarrow$ d $\chi_{1}^0$ )&     2.37791027E-02&  2.37748353E-02  \\
 BR($\tilde{d}_L$ $\rightarrow$ d $\chi_{2}^0$ )&    3.09521294E-01&   3.05370011E-01  \\
 BR($\tilde{d}_L$ $\rightarrow$ d $\chi_{3}^0$ )&    1.67844811E-03&   1.79694702E-03 \\
 BR($\tilde{d}_L$ $\rightarrow$ d $\chi_{4}^0$ )&    1.68190300E-02&   1.80395085E-02 \\
 BR($\tilde{d}_L$ $\rightarrow$ u $\chi_2^{-}$ )&    6.01085498E-01&  6.00388839E-01 \\
 BR($\tilde{d}_L$ $\rightarrow$ u $\chi_2^{-}$ )&    4.71166268E-02&  5.06298586E-02 \\
\hline
\end{tabular}
\caption{$\tilde{d}_L$ decays in the MSSM for SPS-1a, calculated
in both BRIDGE and SUSY-HIT. The decay widths listed are in GeV.}
\label{tab:dldecays}
\end{table}

The results of the comparison between BRI and SUSY-HIT are very
good.  For the decays listed in Tables~\ref{tab:chidecay} and
\ref{tab:dldecays} the differences between BRIDGE and SUSY-HIT are
on the percent level or less.  Examining some of the decays where
loop corrections are important the agreement stays on the order of
5\% difference at most.  We have only compared particle decay
tables for particles that have two body decays only, or just the
partial widths for those particles which have both two and three
body decays.

\subsubsection{Slepton Spin Correlations Example}
As another check of the DGE program, specifically the executable
{\tt runDGEsusy.exe}\footnote{Recall the {\em only} difference
between this and the non-SUSY version of DGE is the input
parameters are fixed.}, we can examine how well it keeps possible
spin correlations.  As an example of spin correlation we will look
at the example studied in~\cite{barr}, in which the process
\begin{equation}
pp>\tilde{l}^{+}\tilde{l}^{-}>l^{+}l^{-}\chi^0_1\chi^0_1
\end{equation}
was studied.  It was shown in~\cite{barr} that by studying the
variable
\begin{equation}
\cos \theta^{*}_{ll}\equiv \tanh^{-1} \left(\frac{\Delta
\eta_{ll}}{2}\right)
\end{equation}
where $ll$ refers to the outgoing $l^{+}$ $l^{-}$ pair, that this
variable encodes the spin correlation of the sleptons that were
originally produced.  Since this spin correlation variable, which
is similar to the variables $\eta_{+}$ and $\eta_{-}$ studied
in~\cite{toppartner}, takes into account opposite sides of the
diagram it is not obvious that decaying the particles separately,
as in DGE, will capture most of the spin dependent effects.

\begin{figure}[!h]
\begin{center}
\includegraphics[width=15cm]{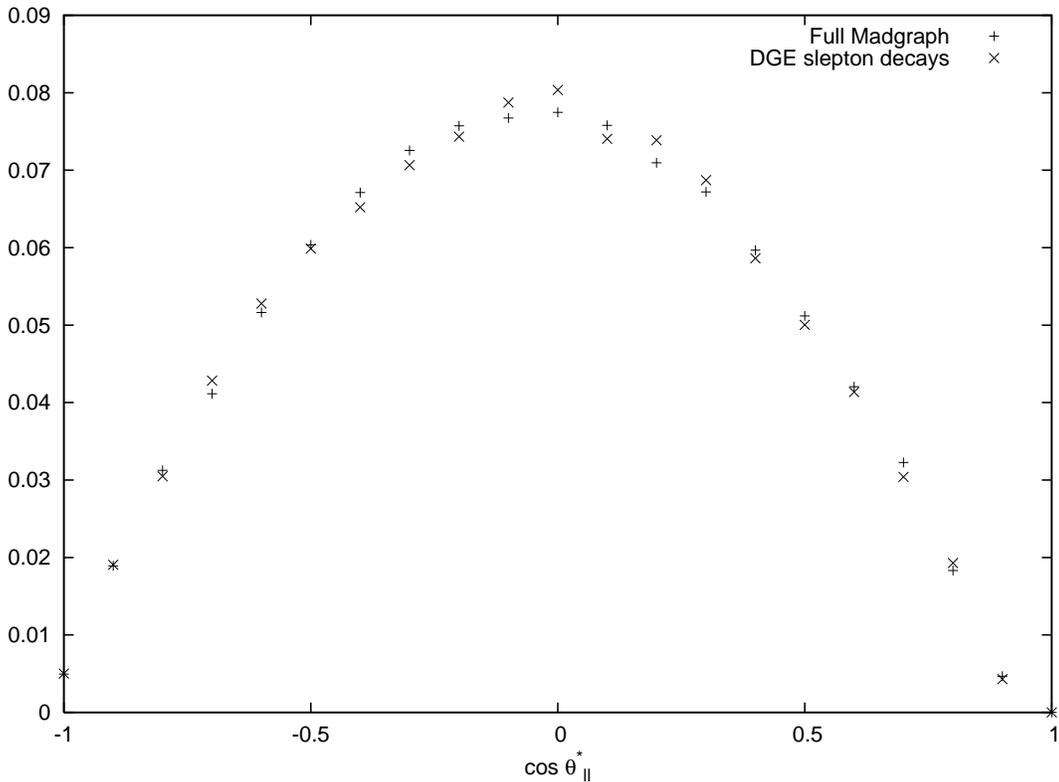}
\end{center}
\caption{Normalized histogram of $\cos \theta^{*}_{ll}$ computed
from MadGraph $\tilde{l}\tilde{l}$ events decayed with BRIDGE
compared to MadGraph $l^{+}l^{-}\chi^0_1\chi^0_1$ events.}
\label{fig:DGEvsMGbarr}
\end{figure}

In Figure~\ref{fig:DGEvsMGbarr} we plot $\cos \theta^{*}_{ll}$ for
SPS-1a from the full matrix element calculation in Madgraph and
compare to the results of first generating
$pp>\tilde{l}^{+}\tilde{l}^{-}$ in Madgraph and then decaying the
sleptons using DGE.  What we find is that DGE seems to keep a
remarkable amount of spin correlation as compared to just
generating a flat matrix element, which for comparison was done
in~\cite{barr}.  Of course this is only a one step decay chain but
nevertheless the results are very encouraging. It would be very
interesting as future work to quantify more fully the amount of
spin correlation that DGE captures in more complicated decay
chains.  In particular studying some of the examples/variables
suggested for spin correlations in the MSSM suggested
in~\cite{toppartner,spin} would serve as a good starting point.

\subsubsection{SPS2}

\begin{table}[!h]
\centering
\begin{tabular}{|l|c|c|}
\hline
$\tilde{g}$ Decays  & BRIDGE $\Gamma$(GeV) & SUSY-HIT $\Gamma$(GeV) \\
\hline
 $\tilde{g} \rightarrow d \bar{d} \chi_{1}^0$ &    2.156 $\times 10^{-5}$ &  3.486 $\times 10^{-5}$  \\
 $\tilde{g} \rightarrow d \bar{d} \chi_{2}^0$ &    0.867 $\times 10^{-4}$ &   1.136 $\times 10^{-4}$  \\
 $\tilde{g} \rightarrow d \bar{d} \chi_{3}^0$ &    2.356 $\times 10^{-7}$ &   1.480 $\times 10^{-7}$ \\
 $\tilde{g} \rightarrow d \bar{d} \chi_{4}^0$ &    3.365 $\times 10^{-6}$ &    5.179 $\times 10^{-6}$ \\
 $\tilde{g} \rightarrow d \bar{u} \chi_1^{+}$ &    1.781 $\times 10^{-4}$ &   2.369 $\times 10^{-4}$ \\
 $\tilde{g} \rightarrow d \bar{u} \chi_2^{+}$ &    6.643 $\times 10^{-6}$ &  9.377 $\times 10^{-6}$ \\
\hline
\end{tabular}
\caption{Some of the $\tilde{g}$ decays in the MSSM for SPS-2, calculated
in both BRIDGE and SUSY-HIT. The decay widths listed are in GeV.}
\label{tab:gludecays}
\end{table}

The SPS2 spectrum has squarks heavier than the gluino, so
it gives a useful check of 3-body decay widths. We do not expect
extremely close agreement with SUSY-HIT, as it includes loop
effects. We present a comparison in Table~\ref{tab:gludecays}.
It is reassuring that the widths are relatively close. (The three-body
code accurately computes the muon decay width, so it seems reasonable
to expect that the differences here are due to the loops.)

\subsubsection{Gluino distributions}

\begin{figure}[!h]
\begin{center}
\includegraphics[angle=270,width=11cm]{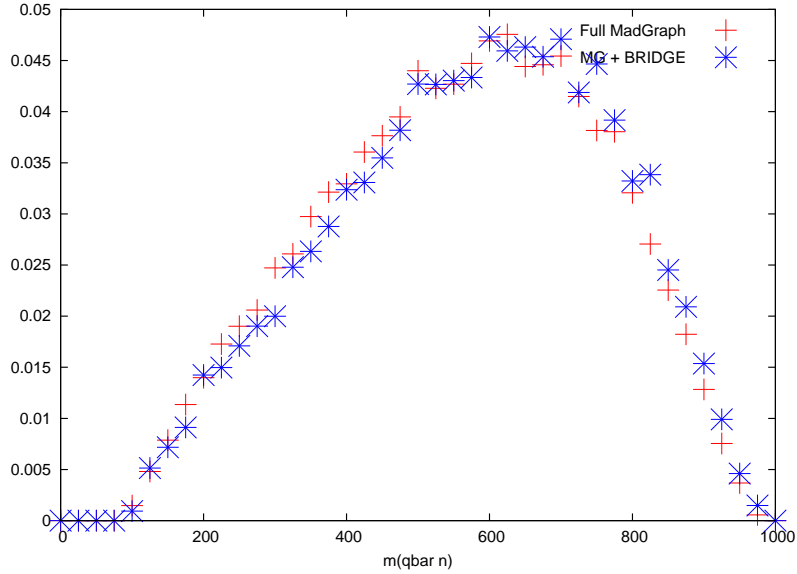}
\end{center}
\caption{Normalized histogram of $m(\bar{q}\tilde{\chi}_0^1)$ computed
from MadGraph $\tilde{g}\tilde{g}$ events decayed with BRIDGE
compared to MadGraph $\tilde{g}q\bar{q}\tilde{\chi}_0^1$ events.}
\label{fig:gluQBarN}
\end{figure}

\begin{figure}[!h]
\begin{center}
\includegraphics[angle=270,width=11cm]{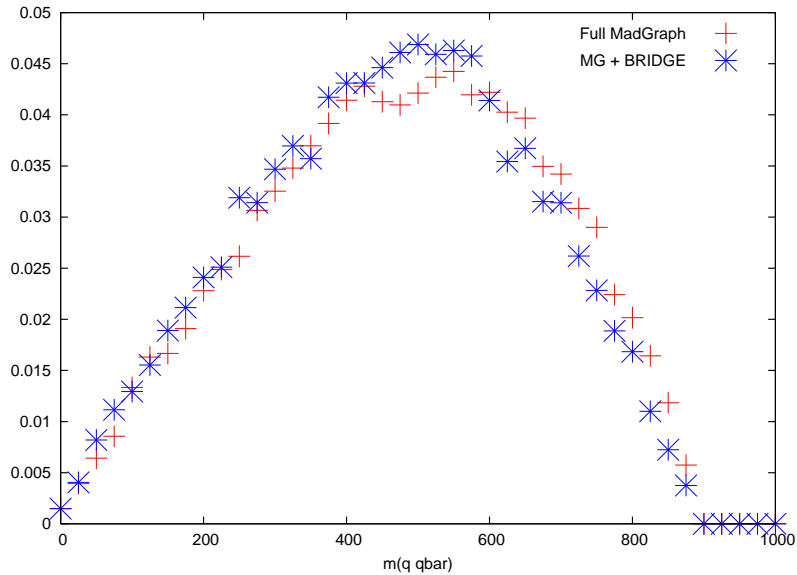}
\end{center}
\caption{Normalized histogram of $m(\bar{q}q)$ computed
from MadGraph $\tilde{g}\tilde{g}$ events decayed with BRIDGE
compared to MadGraph $\tilde{g}q\bar{q}\tilde{\chi}_0^1$ events.}
\label{fig:gluQQBar}
\end{figure}

As a further check of the 3-body code, we implemented a toy model with
a gluino at 1 TeV, one squark at 1.025 TeV, and a neutralino at 100 GeV.
We then simulated gluino production and decay with $\tilde{g}
\rightarrow q\bar{q}\tilde{\chi}_0^1$, both fully in MadGraph and MadGraph
$\tilde{g}\tilde{g}$ production with
gluinos decayed by BRIDGE. We plot the resulting distributions in
Figures~\ref{fig:gluQBarN} and \ref{fig:gluQQBar}.

\subsection{Loop example: a user-defined vertex for $h\rightarrow gg$}
\label{sec:Hgg}

If one uses the standard HELAS amplitudes, one will necessarily
miss decays that happen only at loop level. In the Standard Model
these decays are potentially important; for instance, we might
want to consider $h \rightarrow gg$ or $h \rightarrow \gamma
\gamma$. We have provided an example of how to incorporate such
decays in BRIDGE; if one has already implemented such a vertex for
MadGraph, it requires minimal effort to adapt it for BRIDGE. One
implementation of such couplings is present in the {\tt heft} model
included in MadGraph. Here we discuss an alternative approach,
directly implementing the loop diagram.

First, one must implement the desired vertex as a new HELAS
routine. We provide an example, \texttt{vvsjxx.F}, for the
Higgs--glue--glue vertex in the Standard Model (we consider only
the top loop contribution). To calculate the correct momentum
dependence of the vertex one can use a library like FF or
LoopTools to evaluate integrals~\cite{FFLoops}. You will have to
be sure to link your new routine in the HELAS library (or add it
to the makefile for BRIDGE). Next, as in MadGraph, one should add
an additional character \texttt{J} at the end of the relevant line
of \texttt{interactions.dat} for the vertex.\footnote{We
thank Aaron Pierce for explaining the procedure for adding
loop-level vertices in MadGraph.} This character is
stored by the relevant \texttt{Vertex} object of BRIDGE, and then
passed to the corresponding amplitude function (in this case,
\texttt{VVSAmp}) as a variable called \texttt{usrFlag}. Now, the
code in \texttt{Amplitude.cpp} has lines set off by
\texttt{\#ifdef USE\_VVSJXX} and \texttt{\#endif}. If this
preprocessor flag is set, the code declares the new Fortran
routine \texttt{vvsjxx\_}, which is called if the special
character \texttt{J} is passed from the amplitude. More HELAS
routines can always be added in this manner; the only part of the
BRIDGE code that should need modification to deal with them is
\texttt{Amplitude.cpp}, so long as they involve only scalars,
vectors, and spin one-half fermions. (Note that yet another option
is to directly code the one-loop amplitudes in C++, if one is not
also planning to use the new HELAS routine with MadGraph or some
other program.)

For the width $h \rightarrow gg$, we expect at one loop:
\begin{equation}
\Gamma(h \rightarrow gg) = \frac{\alpha m_h}{8 \sin^2 \theta_W}
\left(\frac{m_h}{m_W}\right)^2 \left(\frac{\alpha_s}{3
\pi}\right)^2 \left|I\left(\frac{m_t}{m_h}\right)\right|^2,
\end{equation}
where $I(x)$ is an order-one quantity that can be expressed as an
integral over a Feynman parameter. We compare the numerical result
of the one-loop calculation to the output of BRIDGE for the same
quantity in Table~\ref{tab:Hgg}. The BRIDGE numbers agree within
about 2\%.

% table: h -> gg results compared to expectations
\begin{table}[ht]
\centering
\begin{tabular}{ccc}
Higgs Mass & $\Gamma(h\rightarrow gg)$, 1-loop & $\Gamma(h\rightarrow gg)$, BRIDGE \\
\hline
60 & $2.55 \times 10^{-5}$ & $2.51 \times 10^{-5}$ \\
120 & $2.13 \times 10^{-4}$ & $2.09 \times 10^{-4}$\\
180 & $7.77 \times 10^{-4}$ & $7.64 \times 10^{-4}$ \\
240 & $2.08 \times 10^{-3}$ & $2.05 \times 10^{-3}$ \\
300 & $4.99 \times 10^{-3}$ & $4.88 \times 10^{-3}$ \\
\end{tabular}
\caption{Higgs partial widths to two gluons, one-loop calculation
versus BRIDGE result with one-loop HELAS vertex added. All numbers
are in GeV.} \label{tab:Hgg}
\end{table}

%%%%%%%%%%%%%%%%%%%%%%%%%%%%%%%%%%%%%%%%%%%%%%%%%%%%%%
%%%%%%%%%%%%%%%%%%%%%%%%%%%%%%%%%%%%%%%%%%%%%%%%%%%%%%
\section{Conclusions and Future Development}
\setcounter{equation}{0} \setcounter{footnote}{0}
%%%%%%%%%%%%%%%%%%%%%%%%%%%%%%%%%%%%%%%%%%%%%%%%%%%%%%
%%%%%%%%%%%%%%%%%%%%%%%%%%%%%%%%%%%%%%%%%%%%%%%%%%%%%%

BRIDGE can speed up the process of simulating a new model, when
used with other tools like MadGraph. It also provides a quick
tree-level calculator for decay rates, which can be a useful start
for understanding the phenomenology of a model. Furthermore, we
have seen that BRIDGE accurately models certain nontrivial angular
effects. One remaining issue is to consider finite (but narrow) width effects: currently intermediate
particles in a long decay chain will be exactly on shell, whereas
we would like to have a Breit-Wigner distribution (while keeping
as much angular information as possible).

In addition, we plan to continue working with the MadGraph authors
to ensure that BRIDGE will interact smoothly with their code. It
should also be possible to use BRIDGE independently of MadGraph,
provided you have the HELAS libraries. BRIDGE could be applied to
a wider array of models if new HELAS routines are developed (e.g.,
for particles of spin greater than 1), so this is another
direction in which further work could be useful.  The potential
usefulness of BRIDGE and other intermediate programs for studying
beyond the SM physics also suggests the need for a standard
definition of a new physics model akin to what we have used in the
Madgraph model definitions.  One final direction to consider involves
the variable {\tt SPINUP} in the Les Houches format. We have seen
that we obtain reasonably accurate angular distributions by using the helicity
in the same frame for every decay along a chain; we store this
helicity in the {\tt SPINUP} column. As with any choice of a single
number to characterize spin, this is losing some information.
Given the degree of interest in spin effects in the collider
phenomenology community, perhaps
we should consider expanding the Les Houches format to allow
a more sophisticated presentation of spin information?

The code for BRIDGE is distributed at
\texttt{http://lepp.cornell.edu/public/theory/BRIDGE/}. We
encourage you to report bugs or (even better) fix them or
contribute new features. If you use BRIDGE in a publication we
request that you cite this document. An up-to-date version of this
manual will be maintained on the BRIDGE website.

%%%%%%%%%%%%%%%%%%%%%%%%%%
\section*{Acknowledgments}
We are grateful to the MadGraph development team for their
assistance and their responsiveness to our bug reports and feature
requests. We also thank the users of BRIDGE who have helped
with bug-fixing; an up-to-date list of them can be found on
the BRIDGE website. P.M. would like to thank the Aspen center for its
hospitality while part of this work was completed. M.R. would like
to thank the theory group at Harvard for its hospitality while a
portion of this work was completed. M.R. is supported by a
National Science Foundation Graduate Research Fellowship and in
part by the NSF grant PHY-0139738.

\end{document}